\documentclass[preprint]{emulateapj}

\shortauthors{Zaritsky {\it et al.}}
\shorttitle{Distance Estimators and the FM}

\begin{document}

\title{Testing Distance Estimators with the Fundamental Manifold}

\author{Dennis Zaritsky$^1$, Ann I. Zabludoff$^1$, and Anthony H. Gonzalez$^2$}

\affil{$^1$ Steward Observatory, University of Arizona, 933 North Cherry Avenue, Tucson, AZ 85721, USA}

\affil{$^2$ Department of Astronomy, University of Florida, Gainesville, FL, 32611-2055, USA}

\begin{abstract}
We demonstrate how the Fundamental Manifold (FM) can be used to cross-calibrate distance estimators even when those ``standard candles" are not found in the same galaxy. Such an approach greatly increases the number of distance measurements that can be utilized to check for systematic distance errors and the types of estimators that can be compared. Here we compare distances obtained using SN Ia, Cepheids, surface brightness fluctuations, the luminosity of the tip of the red giant branch, circumnuclear masers, eclipsing binaries, RR Lyrae stars, and the planetary nebulae luminosity functions. We find no significant discrepancies (differences are $< 2 \sigma$) between distance methods, although differences at the $\sim$ 10\% level cannot yet be ruled out.
The potential exists for significant refinement because
the data used here are heterogeneous B-band magnitudes that will soon be supplanted by 
homogeneous, near-IR magnitudes. We illustrate the use of FM distances to 1) revisit the question of the metallicity sensitivity of various estimators, confirming the dependence of SN Ia distances on host galaxy metallicity, and 2) provide an alternative calibration of H$_0$ that replaces the classical ladder approach in the use of extragalactic distance estimators with one that 
utilizes data over a wide range of distances simultaneously.
\end{abstract}

\section{Introduction}

Distance determinations to galaxies are key to many areas of study in extragalactic astronomy, but remain problematic. The potential for systematic errors lingers above much of the work,
particularly at the accuracy level of a percent or less currently desired \citep[see review by][]{freedman}. These concerns can be addressed through cross calibration among techniques that have independent challenges. For example, one estimator might have a hidden dependence on metallicity, another might be more susceptible to internal extinction. One state-of-the-art recent study compares distances obtained using SN Ia and Cepheid variable stars in the same galaxies \citep{riess} to
claim a distance calibration precision of a few percent. 

Such cases, where one has distance measurements from a variety of methods for a
specific galaxy, are rare for two reasons.
First, because distance estimators are often related to distinct stellar populations, they are not often found within the same stellar system.
The classic example is the dichotomy between the Galactic environments of RR Lyrae and Cepheids \citep{baade}. Second, and currently most relevant, is that several factors conspire to limit estimators to distinct distance regimes. Cepheid observations, like those of RR Lyrae, eclipsing binaries and the tip of the red giant branch (TRGB), require observations that resolve the stellar populations of the system and are therefore limited to relatively
nearby galaxies, even when using the high angular resolution provided by the {\sl Hubble Space Telescope}. On the other hand, SN Ia, due to their rarity, tend to be found in more distant galaxies simply because of the gain in search volume with increasing distance. These effects make it quite difficult to compare the results among certain classes of distance estimators. The \cite{riess} study cited above, which is among the most extensive and complete, is based on only eight galaxies with both Cepheid and SN Ia measurements.

One solution to this limitation relies on identifying a distance estimator that can be applied
to {\sl all} galaxies, near and far, early or late type, giant or dwarf. Even if such an estimator has a large intrinsic 
distance uncertainty for any individual galaxy, and so is not ultimately superior to the other estimators, the gains reaped by being able to compare the relative precision and accuracy of the preferred estimators over many tens, if not hundreds of galaxies, is critical. Distance determinations from 
scaling relations, such as Tully-Fisher \citep{tully} or Fundamental Plane \citep{d87, dd87}, 
span the range of distances probed by many methods and so
provide a fiducial against which different distance estimators can be compared. 
Their shortcoming is that they are each applicable to different, narrow classes of galaxies, limited in type and luminosity, and therefore do not resolve the entire problem. 

In a series of papers, we have presented a new scaling relation that is applicable to {\sl all} galaxies, regardless of luminosity or morphological class \citep{zgz, zgza, zzg, zzgb}. This relation, termed the Fundamental Manifold (FM) in reference to its most direct antecedent, the Fundamental Plane,  presents an opportunity to uncover hidden systematic differences among the current set of distance estimators. Here we use existing data, distances from the NED 1-D database and photometry and kinematics compiled within the Hyperleda database \citep{hyperleda}, to test this
basic idea. Neither the data, due to their inhomogeneity, nor the sample, due to the intractable selection criteria, are optimal for precise cross-calibration of distance estimators and eventual application to related
problems, such as the measurement of H$_0$. Therefore, the results presented here should be viewed as only illustrative of what an optimized treatment using the FM could yield.
In \S2 we discuss our selection of the sample, in \S3 we present the results of this comparison
among distance estimators, illustrate how the FM can be used to trace potential internal systematic dependencies, and reverse the calibration and use the FM to obtain distances and estimate $H_0$. In \S4 we summarize our results.

\begin{figure}[]
\begin{center}
\plotone{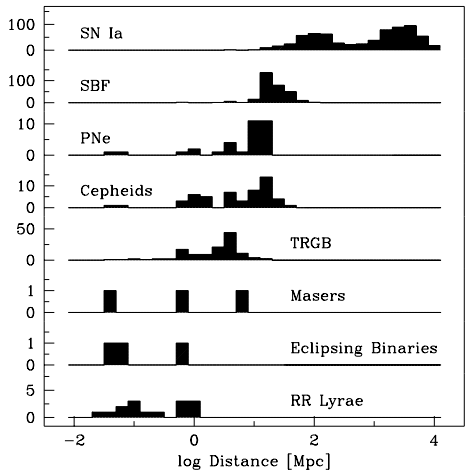}
\caption{Distribution of distances as a function of distance-measuring technique among NED-1D sample (from top, supernova Ia peak luminosity corrected for decay rate, planetary nebular luminosity function, surface brightness fluctuations, Cepheid period-luminosity relation, the luminosity of the tip of the red giant branch, circumnuclear masers, eclipsing binaries, and the RR Lyrae period-luminosity relation). Y-axis arbitrarily normalized to 
match scales among subsamples. Although some distance estimators span the same range of distances, the differences among distance ranges can be as large as two orders of magnitude.}
\label{fig:distances}
\end{center}
\end{figure}

\section{The Data}

\subsection{Distances}

We utilize the NED 1-D database, which is a compilation of published distances from a wide ranging set of methods. Without prejudice for or against any of the entries in the database, we calculate the weighted average (weighting inversely by the quoted uncertainty and neglecting the measurement if no uncertainty is given) for each of the distance estimators, for each galaxy. We require at least two measurements using a particular method to calculate a mean distance measurement for a given galaxy. There are 1153 galaxies for which we retrieve a distance using at least one of the methods we are comparing. In Figure \ref{fig:distances} we show how distance estimators are often effectively constrained to distinct distance ranges. More critically, the distribution of objects with SN Ia distance measurements overlaps very little with the other methods, limiting the number of systems that can be used to directly check for potential systematic errors in the cosmologically-critical Ia distances. 

\begin{figure}[]
\begin{center}
\plotone{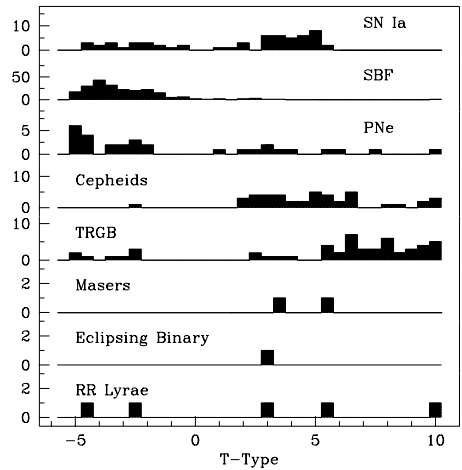}
\caption{Distribution of morphological types as a function of distance measuring technique among NED-1D sample. Y-axis arbitrarily normalized to 
match scales among subsamples. Differences among the types of galaxies accessible to various methods are clear.}
\label{fig:morphs}
\end{center}
\end{figure}

\subsection{Photometry and Kinematics}

To obtain the structural parameters necessary to place a galaxy on the FM, we search 
the Hyperleda database \citep{hyperleda}, which provides morphologies, kinematic measurements, magnitudes and structural parameters, the latter two based on the work of \cite{prugniel}, for those galaxies for which we obtained distances from NED 1-D.
For the \ion{H}{1} rotation measurements, we adopt the homogenized values provided in Hyperleda, which include an inclination correction. We calculate weighted averages of both the stellar rotation and velocity dispersion measurements where available, and only accept values only if there are at least two measurements of either quantity, and if either the rotation or dispersion is  larger than 10 km sec$^{-1}$, to avoid the systems with the largest  fractional uncertainties. For the morphologies, we average T-Types using uniform weights and require a minimum of only one measurement. The photometric parameters are
taken from the archival B-band data, which is the most prevalent. There are various reasons (star formation, extinction) to use redder-band photometry, and we will discuss results using the available I-band data. In the future, we will use the S$^4$G database \citep{sheth} to obtain photometric parameters at 3.6$\mu$m, which appear to significantly reduce the scatter in scaling relations \citep{freedman11, freedman11a}.  The distribution of morphologies, for the subsample of galaxies for which morphologies are catalogued, is shown in Figure \ref{fig:morphs}. Again, we find that certain distance estimators are limited in their coverage, making comparison on an object-by-object basis difficult. There are 343 galaxies with both distance measurements and the necessary data
to place them on the FM.

For a subset of the data, we also obtain an optical color, $(B - V)_0$, 
from the RC3 catalog \citep{dev}
to investigate the role of color, and, by inference, of differences in the stellar mass-to-light ratio, on
a galaxy's deviation from the FM. We use
observed magnitudes rather than k-corrected rest-frame magnitudes,
because, despite the large range in distance of the NED 1-D sample, those galaxies with {\sl all} of the necessary data
lie at $cz <$ 6000 km sec$^{-1}$. One evident avenue for progress is
obtaining these data for a larger fraction of the galaxies with SN Ia distances, and thereby
also extending the sample to greater distances. 

Our compilation of the available data is presented in Table \ref{tab:data} (complete Table available electronically). 

\begin{deluxetable*}{lrrrrrrrrrrrrrrrr}
\tablecaption{The Sample}
\tablewidth{0pt}
\tablehead{
\colhead{Name}&
\colhead{B$_T$} &
\colhead{(B-V)$_0$} &
\colhead{$\mu_e$} &
\colhead{2r$_e$} &
\colhead{v$_{r,HI}$} &
\colhead{v$_{r,*}$} &
\colhead{$\sigma$} &
\colhead{D$_{C}$} &
\colhead{D$_{T}$} &
\colhead{D$_{Ia}$} &
\colhead{D$_{SBF}$} &
\colhead{D$_{PNe}$} &
\colhead{D$_{RR}$} &
\colhead{D$_{M}$} &
\colhead{D$_{Bi}$} &
\colhead{$cz$} 
 \\
&&&&
[$\prime\prime$]&
[km/s]&
[km/s]&
[km/s]&
[Mpc] &
[Mpc] &
[Mpc] &
[Mpc] &
[Mpc] &
[Mpc] &
[Mpc] &
[Mpc] &
[km/s]
\\
}
\startdata
NGC7814&11.38&0.83&22.23&118.1&230.7&0.0&172.9&...&...&...&12.70&...&...&...&...&1047\\
NGC0055&8.68&0.54&22.18&400.1&58.9&0.0&0.0&1.92&2.10&...&...&...&...&...&...&142\\
NGC0147&10.65&0.78&24.02&376.8&0.0&0.0&26.4&...&0.64&...&...&...&0.63&...&...&$-$176\\
NGC0150&11.90&0.50&21.40&63.4&164.9&0.0&0.0&...&...&...&...&...&...&...&...&1566\\
NGC0185&10.38&0.73&22.11&177.1&0.0&0.0&23.3&...&0.61&...&...&...&...&...&...&$-$201\\
NGC0205&9.00&0.82&21.82&292.5&0.0&22.0&33.5&...&0.80&...&...&...&...&...&...&$-$258\\
NGC0221&9.19&0.88&18.75&65.2&0.0&37.8&80.9&...&0.69&...&0.77&...&0.82&...&...&$-$203\\
NGC0224&4.88&0.68&19.89&802.0&0.0&75.1&183.2&0.80&0.79&...&0.75&0.72&0.77&...&0.75&$-$294\\
NGC0247&9.42&0.54&23.37&492.2&249.8&0.0&0.0&3.76&3.60&...&...&...&...&...&...&142\\
NGC0300&8.51&0.58&22.87&594.5&85.1&0.0&0.0&1.99&1.96&...&...&...&...&...&...&150\\
\enddata
\label{tab:data}
\end{deluxetable*}

\section{The Fundamental Manifold and Tests of Distance Estimators}

\subsection{Overview of the Fundamental Manifold}

The FM has been described in detail elsewhere \citep{zzg,zzgb}, but, in summary, it is
derived from the virial theorem with the additional information that the mass-to-light ratio within
$r_e$, $\Upsilon_e$, can be either 1) measured independently (e.g., via strong gravitational lensing) or 2) estimated using an empirical function of the internal kinematics, $V$, and the surface brightness within $r_e$, $I_e$. In the latter case, we have found that the structural differences (spatial and kinematic) among galaxies, other than those captured in the variations in $\Upsilon_e$, are insignificant given the observational errors ($\sim 0.1$ dex). The FM is 
a conceptually simpler scaling relation than either Tully-Fisher (TF) or Fundamental Plane (FP) in that
it can be expressed independent of wavelength and has only a zero point term to calibrate (rather than slope in TF or ``tilt" in FP):
$$\log r_e = 2 \log V - \log I_e - \log \Upsilon_e - C, \eqno{(1)}$$
where $V$ is expressed as the combination of pressure ($\sigma$) and rotational support ($v_r$), $V^2 \equiv \sigma^2  + v_r^2/\alpha$ \citep[$\alpha$ is expected to lie between 2 and 3;][]{weiner,zzgb}), and $C$ is a constant that is effectively the zero point of the FM distance estimator. With measurements (or estimates) of all terms on the right hand side of the equation, one solves for $r_e$, which, in concert with the angular measurement of $r_e$, provides the distance. Because we generally lack independent estimates of $\Upsilon_e$, we use the empirical relationship for $\Upsilon_e(V,I_e)$ derived from existing data \citep{zzgb}. Use of the empirical relationship reintroduces wavelength dependences and additional complexity (analogous to FP ``tilt"), but the FM still retains the advantage over previous scaling relations in that it is applicable to all galaxies.

As with any empirical relationship, the relationship changes slightly depending on the data
used to define it, and uncertainties in the fitting function ultimately translate to  systematic uncertainties in the distances estimated using 
the FM. Consequently, comparing FM distances to those obtained using other distance estimators provides a test of both the FM and the independent distance estimators. Comparisons of the FM and multiple independent methods will allow us to distinguish between systematic problems in the FM versus problems with the inditvidual independent estimators. 
Here we choose to define the fitting function using only the spheroidal galaxies 
in the original data \citep{zzg}, which
are expected to have minimal stellar mass-to-light ratio variations. We exclude globular
clusters and ultracompact dwarfs, whose nature is still controversial \citep[see][]{zzgb} and that extend the
FM beyond the $r_e$ range needed for this work.
Although changes in the fitting function for $\Upsilon_e$ will affect the results, the fitting function is most
constrained in exactly the parameter range most relevant here --- that of disk and spheroid galaxies
with luminosities around $L_*$. 
The functional form we use is
\begin{eqnarray}
\log \Upsilon_e &=& 2.12 - 0.28 \log V - 0.82 \log I_e + 0.15 \log ^2 I_e\\
&&  + 0.26 \log ^2V - 0.09 \log V \log I_e.
\end{eqnarray}
This equation provides an estimate of $\Upsilon_e$ in the I-band. There are two related, but distinct, questions
that will be addressed in the following section: how much does the use of B-band magnitudes  
increase the scatter (there would be no increase if all galaxies had the same $B-I$ color) and 
how much scatter is introduced in the $I$-band $\Upsilon_e$ by differences in stellar populations?

To test each distance method, we use the distances provided by NED 1-D to calculate the physical value of $r_e$ and then place the galaxy on the FM. In our original work, from which we
take our fitting function for $\Upsilon_e$, we used simple Hubble flow distances with an adopted
Hubble parameter.
Any difference between the true Hubble parameter and that adopted previously will affect the calibration of $\Upsilon_e$ and $C$, manifesting itself as a zero point shift ({\sl i.e.}, as a change in $C$). We therefore have freedom to adjust the zero
point of the FM as needed (see below for our treatment of this) and so our discussion here focuses
on relative differences between distance estimators rather than on an absolute calibration. 

\begin{figure}[]
\begin{center}
\plotone{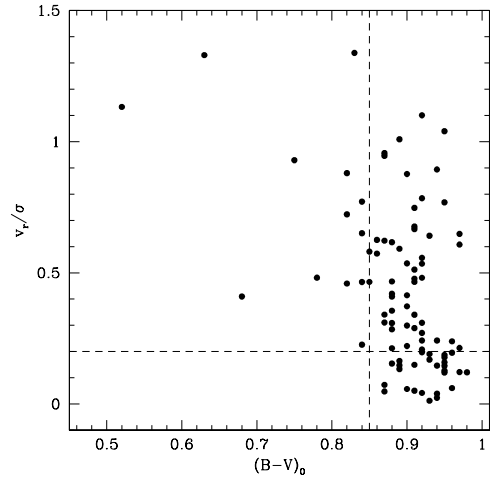}
\caption{The relationship between color and $v_r/\sigma$ for galaxies in our sample that
have published measurements for stellar velocity dispersion,  $\sigma$, and rotation, $v_r$. 
We select the subsample with $B-V > 0.85$ and $v_r/\sigma > 0.2$, separated with dashed lines, for subsequent analysis because it is large and has no
measurable relationship between color and $v_r/\sigma$ (see text for an expanded discussion of this selection). }
\label{fig:colsel}
\end{center}
\end{figure}

\subsection{Morphology Related Differences}

One consideration before proceeding with our comparison of distance estimators and the FM is that of the appropriate choice of $\alpha$ in the definition of $V$. In previous
work we adopted $\alpha = 2$, as also done by others \citep[e.g.,][]{burstein,kassin}. However, reasonable models suggest
that  values between 2 and 3 are plausible \citep[see][]{weiner}. Because $v_r/\sigma$ maps
onto morphological type, and morphological type maps onto color, which is presumably related to 
the stellar mass-to-light ratio,  there is the potential to 
misidentify a relationship between the FM residual (the deviation of a galaxy from the FM) and $\alpha$, with one that is truly due to color (or vice-versa).  Without the proper choice of 
$\alpha$ and some means of avoiding differences in stellar mass-to-light ratios beyond those included in the empirical calibration of $\Upsilon_e$, there will be systematic differences
when comparing distances to early and late type galaxies. We will mitigate this difficulty by 1) proceeding
to optimize our choice of $\alpha$, 2) removing any residual correlation between FM residuals
and color, and 3) providing comparisons among galaxies of similar morphological type.

There is indeed a strong
correlation between $(B-V)_0$ and $v_r/\sigma$ (a Spearman rank correlation test for our sample puts the probability of it arising at random at $4\times10^{-5}$). As such, for the purpose of 
constraining $\alpha$ we work with a subsample
for which $v_r/\sigma$ is independent of color (Figure \ref{fig:colsel}); $(B-V)_0 > 0.85$ and $v_r/\sigma > 0.2$ (the probability of the correlation arising
randomly for this sample is 0.79). This particular part of the analysis, the derivation of $\alpha$, is restricted to early type galaxies because $\sigma$ is generally not available for disk galaxies. As a result, the galaxies in Figure \ref{fig:colsel} are naturally restricted to fairly red colors. If a hidden correlation remains, we expect some of that to influence the
best fit value of $\alpha$, again because $v/\sigma$ is expected to correlate with color, and eventually to lead potentially to differences between FM residuals and morphological type. 

\begin{figure}[]
\begin{center}
\plotone{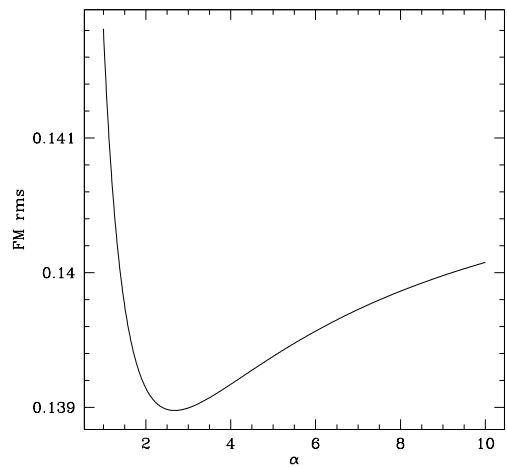}
\caption{Residual about the FM as a function of adopted $\alpha$. The minimum value at 2.68 is chosen for the remainder of the study.}
\label{fig:minalpha}
\end{center}
\end{figure}

For this subsample, we calculate the root mean square (rms) residual about the FM 
(allowing the zero point to 
float) and identify the value of $\alpha$ that minimizes the rms ($\alpha = 2.68$; see Figure \ref{fig:minalpha}). Reassuringly, this value lies within the theoretically plausible range and 
we adopt this value of $\alpha$ for the remainder of this study. However, this calculation is unfortunately quite sensitive to the selected subsample. For example, if we
select galaxies that have $(B -V)_0 > 0.8$ (instead of 0.85), then the minimum rms residual occurs at $\alpha = 1.81$ (although a correlation between color and $v_r/\sigma$ still exists at $> 1\sigma$ confidence for this sample and so this sample is not preferred). The robustness of the $\alpha$ determination needs to be improved both by obtaining a larger sample of galaxies and by performing this analysis with photometry in a filter band that is less sensitive to morphological type (and, by relation, to color). The effect of varying $\alpha$ is principally to alter the relative
offset between early and late type galaxies relative to the FM (see \S4), but does not affect the conclusions presented in \S5.

\begin{figure}[]
\begin{center}
\plotone{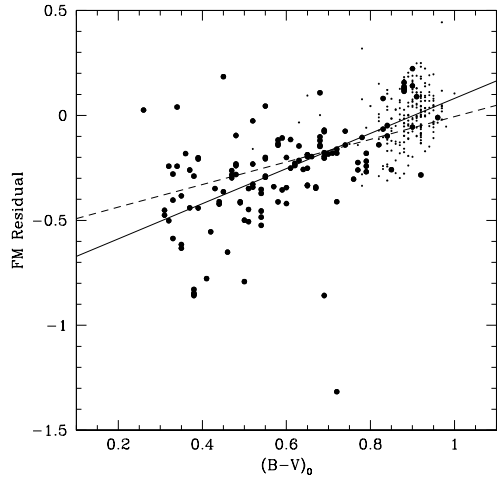}
\caption{FM residual for individual galaxies as a function of color. Pressure supported galaxies (those with measured $\sigma$) are plotted as dots and represent the results using the distances
from surface brightness fluctuations. For the rotationally supported galaxies, plotted as filled circles, we show the results using distances derived from Cepheid measurements, tip of the red giant branch,  SNe Ia, planetary nebulae, and surface brightness fluctuations. The solid line represents the best fit and is used for the color-dependent correction. The dashed line represents the expected color dependence, normalized to cross our fit at $(B-V)_0 = 0.7$, derived from a synthetic stellar population (see text for details).}
\label{fig:color}
\end{center}
\end{figure}

Any remaining systematic behavior of the FM residuals will point to new phenomenon, including
systematic errors in distance estimators.
We begin by examining the dependence of FM residual to color (Figure \ref{fig:color}). The
data show a trend of residual with color. It is not surprising that the 
early-types have near zero mean residual, because the FM relationship we use was defined
using spheroidal galaxies (from a different sample based on a range of sources, see \citet{zgz}, 
but early-types nonetheless). Neither the sense nor the magnitude of the
trend is necessarily straightforward to explain. There is an expectation of a trend with color because the 
stellar mass-to-light ratio will change with color. However, other factors change as 
well. For example, the gas mass fraction of a galaxy correlates with color 
and that mass is unaccounted for in the empirical derivation of $\Upsilon_e$ because
we used only spheroidal systems. 

We use the mean relationship, shown
as the line, to correct all of our data (and so subsequently restrict ourselves to systems with available color measurements). If a particular estimator (including the FM) has a systematic error that depends on galaxy type (color), it will contaminate this plot. As with many of
the issues raised, we expect this problem to be 
mitigated using redder photometry, but proceed nonetheless with the data currently available.  Recall that our expression for $\Upsilon_e$ provides an
estimate of the total $I-$band mass-to-light ratio within $r_e$, even though we are using
$B-$band magnitudes. This means that there are two potential sources of scatter introduced
by stellar population differences. First, there is the possibility of large variations in $B-I$ within
the sample that then introduce scatter because we implicitly assume a constant $B-I$ color (solar) when converting $B$-band luminosity into the corresponding mass in solar masses. 
Second, scatter can arise from stellar population differences and the associated variations in the stellar mass-to-light ratio.

\begin{figure}[]
\begin{center}
\plotone{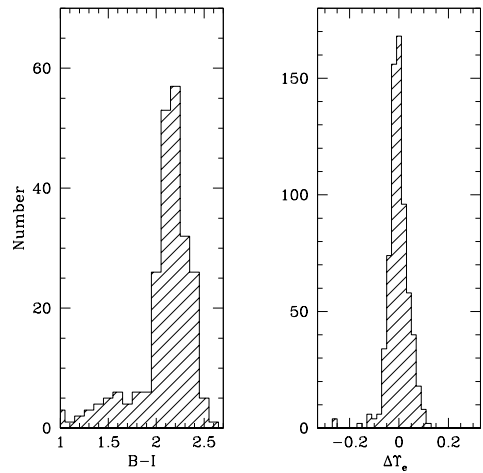}
\caption{The effect of $B-I$ color variations on the scatter in the FM. Left panel shows the distribution of $B-I$ for the galaxies in our sample. The right panel shows the resulting differences in estimates of $\Upsilon_e$ using the $B$ and $I$ band magnitudes, after removing for $B-V$ color dependence. This comparison, which is unphysical, highlights the scatter introduces solely by assuming a 
uniform $B-I$ color for all galaxies, but not assuming that the stellar populations are all the same.
We conclude that the use of $B$-band magnitudes is not a major source of scatter in the FM, if one does correct for differences in stellar mass-to-light ratios using colors.}
\label{fig:bmi}
\end{center}
\end{figure}

We
show, in Figure \ref{fig:bmi}, the distribution of $B-I$ colors for our galaxy sample, and note that a large fraction of the galaxies are fairly
well constrained to colors that vary by about 0.3 mag (a constant difference from solar color is not a cause for concern because we eventually renormalize the entire relation using the SBF distances). More importantly, this range of colors
produces small scatter (0.046 dex) in $\Upsilon_e$ between the values
calculated using the $B$-band magnitudes (and the color correction described above)
and the $I$-band magnitudes (with a color correction derived for the $I$-band).
We conclude that variations in $B-I$ colors, beyond those that we remove using a mean
trend of FM residual with $(B-V)_0$, are a minor contributor to the scatter we observe
in the FM ($> 0.1$ dex for these data).   

Our $(B-V)_0$ color correction  
is intended to empirically remove such a correlation, 
although as hinted at previously, this correction
may be removing other factors as well that work in parallel to stellar population differences. In
Figure \ref{fig:color} we plot the expected variation in stellar mass-to-light ratio vs. $B-V$
from \cite{bell}, selecting one particular model (closed box, scaled Salpeter IMF), demonstrating
that the effect we are removing is indeed of the same scale as that expected from stellar 
population variations.

Finally, we estimate the degree to which going to even redder passbands might 
reduce the scatter, when one does not have a color measurement. For the same stellar population
model as in Figure \ref{fig:color}, the slope in the logarithmic relation between $B-V$  and the stellar mass-to-light ratio at $K$ is about half that in $I$. Therefore, if the scatter
in the FM is dominated by stellar population differences, and we do not have colors, we still expect to reduce the scatter by two simply by observing at $K$. Further gains are possible if extinction is important \citep{freedman11a}. If other 
causes, not related to the photometric passband, contribute comparably to the scatter, or if we can partially account for stellar population differences, then we expect less than a factor of two improvement by going into the infrared. For example, the scatter in our $B$-band
FM is the same as in our $I$-band FM ($\sim 0.1$ dex), because we have accounted for the
stellar population differences at $B$.

\begin{figure*}[]
\begin{center}
\plotone{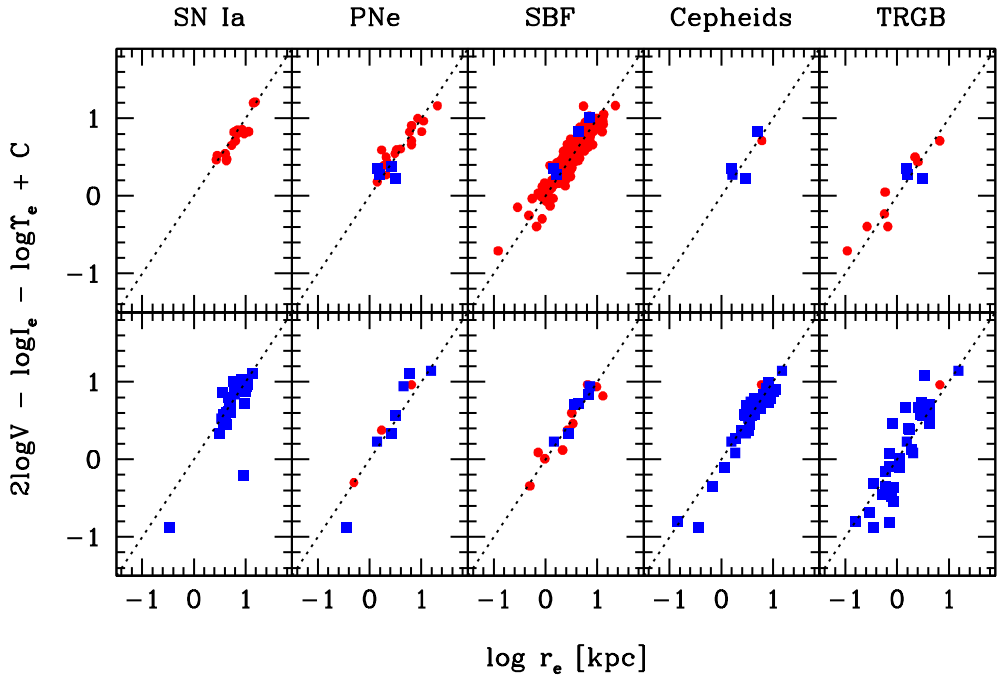}
\caption{FM using measured distances. We compare the FM's using different distance estimators and divide the sample into pressure supported (upper panels) and rotationally supported  (lower panels). Furthermore, we color and shape code according to galaxy morphology (blue squares for late type, red circles for early type).
The x axis is log $r_e$ in kpc, and the y-axis is $2 \log V - \log I_e - \log \Upsilon_e - C$, where $V$ is in km sec$^{-1}$, $I_e$ is in $L_\odot/$sq. pc, and $\Upsilon_e$ is in solar units. $C$ is obtained by calibrating to the sample of surface brightness fluctuation distances (SBF) for the pressure supported galaxies. The line is the 1:1 expectation, not a fit to the data. Individual uncertainties are not plotted, but the scatter is used to estimate the uncertainty in the mean offsets from the 1:1 line.}
\label{fig:comp1}
\end{center}
\end{figure*}

The key to
interpreting our results is to determine whether multiple estimators show the same systematic behavior in FM residuals, in which 
case the problem is most likely to lie with the FM or the various corrections we have implemented, or if the
FM residuals are peculiar for specific estimators, in which case the fault lies with that particular estimator. On the other hand, if none of the estimators exhibit
systematic problems relative to the FM, then we will conclude that the estimators and the FM are accurate distance
estimators to at least within the quoted uncertainties. 

\subsection{Comparing Distance Estimators and the FM}

Our aim is to test a variety of popular distance estimators to determine which, if any, may
lead to distance estimates with systematic errors. Using the FM as a reference, we now compare
estimators that we could not previously test against each other. To do this,
we plot the galaxies on the FM using the mean distances drawn from the NED 1-D data (Figure \ref{fig:comp1}). 
Normalization or slope errors for any particular estimator will suggest that there is a problem
with that estimator. On the other hand, if the normalization or slope errors appear endemic, then the problem will
lie with the FM.

\begin{figure}[]
\begin{center}
\plotone{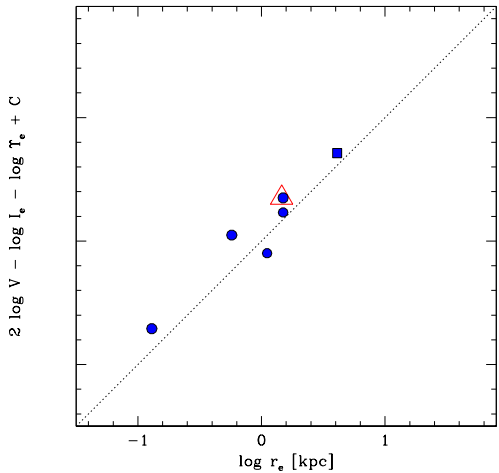}
\caption{FM using measured distances. Same as Figure \ref{fig:comp1}, except we include other distance estimators (maser (blue square), eclipsing binaries (red triangle, plotted large to enhance visibility),  and RR Lyrae (blue circles)). The value of $C$ is as in Figure \ref{fig:comp1}.}
\label{fig:comp2}
\end{center}
\end{figure}

For each distance estimator with more than 10 galaxies with all the necessary data, we present the results, showing 
the pressure supported systems, which have velocity dispersions and sometimes also have stellar rotation values, $v_r$(upper panels in the Figure), and the
purely rotationally supported galaxies, which have no quoted stellar velocity dispersion and for which we adopt the \ion{H}{1} width as a measure of $v_r$ (lower panels). In general, if there is a stellar velocity dispersion measurement, the galaxy is a pressure supported systems ($v_r/\sigma < 1$).  Within each panel we also color-code based on morphology, dividing the early and late type galaxies at a T-Type of 1. We have set the zero point of the FM, $C$, using the results from the pressure-supported, SBF sample, which is our largest subsample of galaxies and also predominantly consists of early-type galaxies, which will
be less susceptible to extinction and stellar population variations.

A cursory examination of the panels shows that there are no disastrous problems with
any of the distance estimators as applied to any of the galaxy subsamples, even though some perform better than others (either in terms of zero point or scatter). There are a few individual galaxy outliers, although it is often the same galaxy that is an outlier in multiple panels because distances are available from multiple estimators, suggesting that fault lies not with 
the distance estimate but rather with one of the other parameters that enters the FM. 
NGC 4704, which has an unusually large residual in the SN Ia panel, is removed from subsequent analysis of the SN Ia distance estimator.
In Figure \ref{fig:comp2} we show the results for galaxies that have distances measured using eclipsing binaries, RR Lyrae, and masers. 

\begin{figure}[]
\begin{center}
\plotone{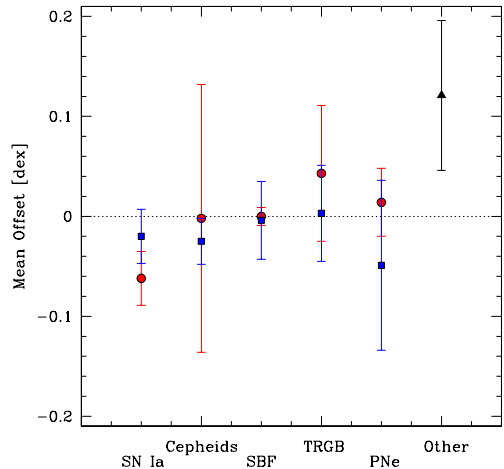}
\caption{Mean offsets from FM using distances derived from different estimators. Red circles denote the results for pressure supported galaxies, while blue squares those for rotation supported galaxies. ``Other" includes results from eclipsing binaries, RR Lyrae and masers for both rotation and pressure supported galaxies. Error bars represent the dispersion in the mean value. The distances are referenced by definition to the surface brightness fluctuation (SBF) calibration for pressure supported galaxies.}
\label{fig:comp3}
\end{center}
\end{figure}

In Figure \ref{fig:comp3} we compare the mean
residual and the uncertainty in the mean between the various distance methods, for pressure
and rotationally supported galaxies (the values are given in Table \ref{tab:comp}). The agreement is strikingly good, with the possible
exception of results obtained using the ``other" methods, although there the uncertainty is large because of the small number of galaxies
in that category. None of the differences are statistically significant, but the magnitude of differences is in the range that is potentially important for distance estimators ($\sim$ few to tens of \%). For example, the difference in zero points obtained using the TRGB and
SN Ia methods and pressure supported galaxies is 0.062,  which corresponds to an inferred distance difference of 15\%. 
Unfortunately, for the purpose of identifying if there are any real discrepancies at the target level of a few percent, the error bars
are still somewhat too large due to the large amount of scatter in the FM (arising from the heterogeneous B-band data) and small sample size (as many galaxies lack the full set
of measurements needed to calculate the FM). 

\begin{deluxetable}{lrrl}
\tabletypesize{\scriptsize}
\tablecaption{Mean Offsets from FM}
\tablewidth{0pt}
\tablehead{
\colhead{Distance Estimator} &
\colhead{Pressure Supported} &
\colhead{Rotationally Supported} \\
}

\startdata
Cepheids     & $-0.002 \pm 0.134$ & $-0.025 \pm 0.023$\\
SBF      & $0.000 \pm 0.009$ & $-0.004 \pm 0.039$\\
SN Ia      & $-0.062 \pm 0.027$ & $-0.020 \pm 0.027$\\
TRGB      & $0.043 \pm 0.068$ & $0.003 \pm 0.048$\\ 
PNe & $0.014 \pm 0.034$ & $-0.049 \pm 0.085$\\
\enddata
\label{tab:comp}
\end{deluxetable}

The uncertainties in our
comparison of distance estimators are currently dominated by the number of galaxies in each class and the intrinsic
scatter in the FM. The small sample sizes can be addressed relatively easily. Many galaxies with extant distance measurements, particularly those with SN Ia distances, lack the additional measurements (typically $V$)
necessary to place them on the FM. These can be obtained with ground-based observing. The FM scatter can be improved by going to redder pass-bands. If all of the data had uncertainties similar to the pressure-supported, SBF measurements, we would easily confirm or refute the level of discrepancy among distance estimators found here.

\begin{figure}[]
\begin{center}
\plotone{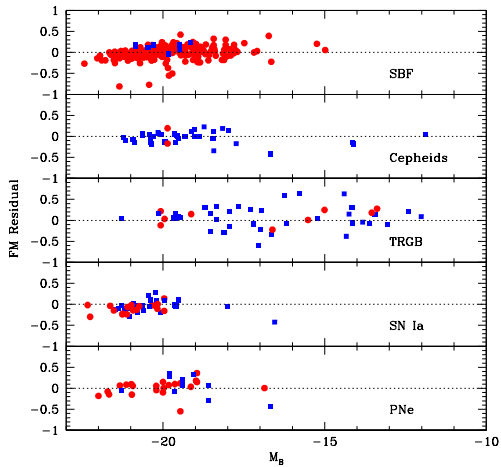}
\caption{Residuals from the FM as a function of galaxy absolute magnitude for different distance estimators. As in other Figures, late type galaxies plotted as blue squares and early type galaxies as red circles. Systematic deviations, if present, suggest a possible abundance dependence in the respective distance estimator due to the correlation between galaxy luminosity and mean chemical abundance.}
\label{fig:abundances}
\end{center}
\end{figure}

\subsection{Testing the Metallicity Dependence of Distance Estimators}

The FM is not only of use in uncovering systematic differences among distance estimators, but is
also potentially useful in uncovering systematic errors in any particular distance estimators within a set of galaxies. One long-running concern is the role chemical abundance might play in affecting distance estimators \cite[see][for discussions of the effects on Cepheids and SNe Ia, respectively]{freedman11,gallagher}. Given the correlation between host galaxy gas-phase metallicity and luminosity \citep{zkh},
we use our sample to examine whether there are potential dependencies on metallicity by measuring correlations between FM residuals and luminosity (a correlation could also implicate another parameter that correlates with luminosity, e.g., mean stellar age). In Figure \ref{fig:abundances} we show the FM residuals as a function of galaxy absolute magnitude for our five
primary estimators. Visually, two of the panels appear to show a trend of increasing FM residual with decreasing luminosity (SBF and SN Ia). However, this similarity among panels overstates the concordance because some of the galaxies, particularly at the bright end, appear in multiple panels. In Table \ref{tab:metallicity}, we present measurements of the relationship between the FM residuals and absolute magnitude for each distance estimator. We list both the fit obtained  using all of the galaxies and after trimming the poorly populated faint end of the galaxy distribution (either at $M_B = -17$ or $-18$). The only two significant correlations are provided by the SBF and SN Ia data (their rank correlation probabilities of arising at random are both $< 0.3\%$, if we confine the
SNe sample to galaxies with $M_B < -18$).  

\begin{deluxetable}{lrrl}
\tabletypesize{\scriptsize}
\tablecaption{Magnitude Dependent Offsets from FM}
\tablewidth{0pt}
\tablehead{
\colhead{Distance Estimator} &
\colhead{All Data} &
\colhead{Faint Tail Trimmed} \\
}

\startdata
Cepheids     & $-0.010 \pm 0.011$ & $0.023 \pm 0.022$\\
SBF      & $0.033 \pm 0.008$ & $0.037 \pm 0.008$\\
SN Ia      & $0.013 \pm 0.032$ & $0.093 \pm 0.045$\\
TRGB      & $0.007 \pm 0.015$ & $-0.043 \pm 0.042$\\ 
PNe & $-0.002 \pm 0.029$ & $0.053 \pm 0.035$\\
\enddata
\label{tab:metallicity}
\end{deluxetable} 

If we attribute the slopes to a metallicity dependence in the estimator, we can
quantify the effect and compare to previous determinations. 
Our best fit for the SN Ia in galaxies with $M_B < -18$ implies a distance
offset of 
$-0.41\pm 0.11$ dex/[O/H]  (for the
4.5 mag/[O/H] relation of \cite{zkh}), which agrees with the sense found by \cite{howell} and 
\cite{kelly} in that
one would underestimate the distance to SNe in more metal rich galaxies,
because SNe are intrinsically more luminous
after light curve correction. It is difficult to quantitatively compare our
results to those of \cite{kelly} for various reasons, including their use of stellar masses vs. our use of luminosities and our calibration to gas-phase abundances. However, they find distance modulus variations that correspond to about 0.15 over the 1 order of magnitude in mass. If we assume that the order of magnitude in mass correspond to the same magnitude range in luminosity, then we would predict a distance effect of 0.23 dex over the corresponding range, which in turn results in a change of 
$1.15 \pm 0.30$ in distance modulus. This is a much larger value than that found by \cite{kelly}, but the discrepancy is not significantly dramatic given our large uncertainties. Likewise, our value
expressed in dex/[O/H] is larger than that found by \cite{howell} ($-0.10 \pm 0.07$), but again the discrepancy is $< 3 \sigma$.  For Cepheids, we find no significant correlation, but our measurement ($0.23 \pm 0.22$ dex/mag or $-0.92 \pm 0.88$ dex/[O/H] if the correlation arises from metallicity) can be compared to previous determinations. Given the large
uncertainties, this result agrees with previous values, which generally range between $-0.1$ and $-0.4$ dex/[O/H] \citep{romaniello}. 

\begin{figure}[]
\begin{center}
\plotone{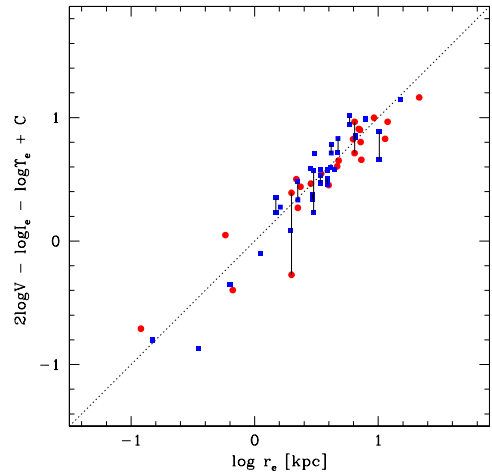}
\caption{The Fundamental Manifold relation for galaxies with distances determined with at least two independent methods. We adopt the average of the distance determinations.
Pressure supported systems are plotted as red circles and rotationally supported systems as blue squares. When we have both measurements of the HI circular velocity and stellar kinematics, we place the galaxy on the FM using each of these and connect the points. These connections illustrate the contribution of the
uncertainties in the kinematics to the scatter.}
\label{fig:FM}
\end{center}
\end{figure}

\subsection{The FM as Distance Estimator}

In Figure \ref{fig:FM} we show the FM for galaxies that have their distances estimated using multiple techniques (we use the average for the adopted distance). The relation is satisfied by galaxies that  range widely over luminosity, morphology, and kinematics.
It is tempting to invert the arguments above and use the FM not as a fiducial against which to test various distance estimators, but as a distance estimator itself. As stressed in the Introduction, these particular data are suboptimal for such a experiment, but we proceed nonetheless .

\begin{figure}[]
\begin{center}
\plotone{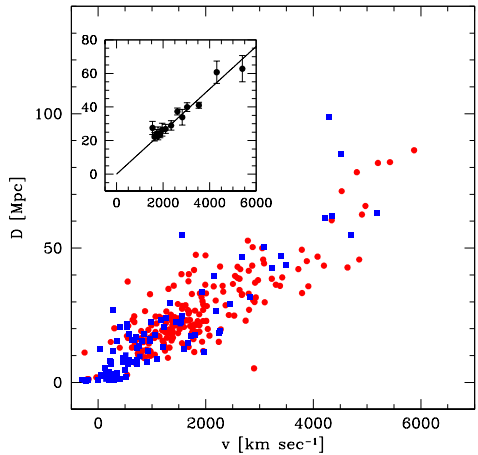}
\caption{Hubble diagram using distances derived using the FM. Pressure supported systems are plotted as red circles and rotationally supported systems as blue squares. Inset shows data at $v > 1500$ km sec$^{-1}$ binned so that each bin contains 10 galaxies. The fitted line is constrained to go through the origin and corresponds to $H_0 = 78 \pm 2$ (random) $\pm 4$ (systematic) km sec$^{-1}$ Mpc$^{-1}$. }
\label{fig:hubble}
\end{center}
\end{figure}

Using the FM now for all of the galaxies in the sample, we estimate a distance to each, and show the relationship between recessional velocity and distance in Figure \ref{fig:hubble}. We define the FM zero point by setting the mean FM residual among estimators to zero, but define that mean in several different ways. First, we use the SBF calibration. Second, we calculate the weighted mean of the residuals in Table 2, excluding our calibration SBF sample. Third, we use the zero point from the SN Ia for pressure supported galaxies. The results differ by at most 4 km sec$^{-1}$ Mpc$^{-1}$ in the inferred $H_0$. We
therefore cite a systematic uncertainty of 4 km sec$^{-1}$ Mpc$^{-1}$ in our $H_0$ measurement below.
We choose to plot the results
using the intermediate value of the FM zero point derived from the three methods (the mean of the residuals of the estimators).
The inset shows the result of fitting the relationship for galaxies with $v > 1500$ km sec$^{-1}$, $3\sigma$ outliers excluded, binned by 10, and forcing the fit through the origin.  We exclude the low $v$ region to minimize the effect of local flows. The resulting slope implies $H_0 = 78 \pm 2$ (random) $\pm$ 4 (systematic) km sec$^{-1}$ Mpc$^{-1}$. The estimate of the systematic uncertainty does not include a variety of potential problems that are ignored here (modeling of bulk flows, internal extinction corrections, adjustment for potential biases in the galaxy sample, etc.). Therefore, while we treat this result cautiously, we present it to show 1) the level of precision possible with
plausible sample sizes and 2) to illustrate how one could determine $H_0$ in a way that utilizes as many (or as few) different distance estimators as desired. 

One of those sources of uncertainty, internal extinction, can be mitigated by going to the infrared, as shown most recently by \cite{freedman11} for {\sl Spitzer} wavelengths where the scatter about the Tully-Fisher relation is
reduced from 0.43 in the extinction-corrected $B-$band data to 0.31 in the non-extinction-corrected
3.6$\mu$m data. We expect analogous gains in the FM for galaxies in which internal extinction is important. Depending on the fraction of galaxies in a sample for which extinction contributes significantly to the observed FM scatter,
proportional gains in the determination of $H_0$ will follow.

\section{Conclusions}

We demonstrate that the fundamental manifold \cite[FM;][]{zzg} provides a fiducial against which any set of distance estimators can be
compared. 
Many, if not most, galaxies will only ever have distances measured with one method, making it
impossible to use them to test distance estimators directly. The FM fiducial, a benchmark for all distance estimators, constrains whether systematic errors exist
among estimators and whether they are creeping into even a single estimator as a function of external parameters, such as metallicity. Large sample size is critical for examining selection effects where, for example, one might want to split the sample in various ways ({\it e.g.,} with environment, luminosity, morphological type, star formation rate). 

Among the distance estimators we examine (SN Ia, Cepheids, surface brightness fluctuations, the luminosity of the tip of the red giant branch, circumnuclear masers, eclipsing binaries, RR Lyrae stars, and the planetary nebulae luminosity functions), we find no statistically significant differences ($> 2 \sigma$), 
but caution that differences that are unacceptably large for state-of-the-art distance uncertainties ($\sim$ few \%) are still allowed by our data. The lack of
precision with which we could discriminate among methods is rooted in sample size, which in certain cases is
still too small, and in the less-than-optimal data available ($B$ band and heterogeneous). Both of
these problems will be resolved with a feasible dedication of resources. 

We illustrate the use of the FM distances to 1) test the metallicity sensitivity of different estimators, confirming the SN Ia distance dependence on host galaxy metallicity and 2) provide an alternative
calibration of $H_0$ that 
replaces the classical ladder approach in the use of extragalactic distance estimators with one that 
utilizes data over a wide range of distances simultaneously.

\begin{acknowledgments}

We thank Barry Madore and Ian Steer for compiling and generously providing the
NED 1-D distance database without which this project could not have been completed.
DZ acknowledges financial support for this work from a
NASA LTSA award NNG05GE82G and NSF grant AST-0307482.
AIZ acknowledges financial support from NASA LTSA award NAG5-11108 and
from NSF grant AST-0206084.
 DZ and AIZ thank the Institute of Astronomy at Cambridge 
University and the Center for Cosmology and Particle Physics at New York University for their
hospitality during the completion of this study.
We acknowledge the usage of the HyperLeda database (http://leda.univ-lyon1.fr)
and thank all those involved for producing this highly useful resource.  This research has made use of the NASA/IPAC Extragalactic Database (NED) which is operated by the Jet Propulsion Laboratory, California Institute of Technology, under contract with the National Aeronautics and Space Administration. We also extend our thanks to those involved in creating and maintaining this excellent resource.

\end{acknowledgments}

\end{document}